# The Moral Foundations of Left-Wing Authoritarianism: On the Character, Cohesion, and Clout of Tribal Equalitarian Discourse

Justin E. Lane[1] Kevin McCaffree[2*] and F. LeRon Shults[3&]

[1] ALAN Analytics S.R.O, Center for Mind and Culture, and Center for Modeling Social Systems, Kristiansand, Norway
[2] Department of Sociology, University of North Texas, Denton, TX, USA
[3] Department of Global Development and Planning, University of Agder, Kristiansand, Norway.

*Corresponding author

Email: leron.shults@uia.no

[&]Authors are listed alphabetically. All authors contributed equally to this work.

**Abstract**

Left-wing authoritarianism remains far less understood than right-wing authoritarianism. We contribute to literature on the former, which typically relies on surveys, using a new social media analytic approach. We use a list of 60 terms to provide an exploratory sketch of the outlines of a political ideology—tribal equalitarianism—with origins in 19$^{th}$ and 20$^{th}$ century social philosophy. We then use analyses of the English Corpus of Google Books ($n > 8$ million books) and scraped unique tweets from Twitter ($n = 202{,}582$) to conduct a series of investigations to discern the extent to which this ideology is cohesive amongst the public, reveals signatures of authoritarianism and has been growing in popularity. Though exploratory, our results provide some evidence of left-wing authoritarianism in two forms: (1) a uniquely conservative signature amongst ostensible liberals using measures derived from Moral Foundations Theory and (2) a substantial prevalence of anger, relative to anxiety or sadness, in tweets analyzed for sentiment. In general, results indicate that this worldview is growing in popularity, is increasingly cohesive, and shows signatures of authoritarianism.

# The Moral Foundations of Left-Wing Authoritarianism: On the Character, Cohesion, and Clout of Tribal Equalitarian Discourse

**Introduction**

In recent years public discourse in the US has become increasingly polarized and sectarian, posing new threats to social stability and democracy (Finkel et al., 2020). Given the seriousness of these societal challenges, it is not surprising that many psychologists, political theorists, and other scholars have attempted to explain the mechanisms underlying them. One of the most influential theories informing this research has been based on the work of Theodore Adorno and colleagues on *The Authoritarian Personality* (Adorno et al., 2019), originally published in 1950. This approach identified authoritarianism with right-wing ideologies and behaviors. Up until recently, right-wing authoritarianism (RWA) was the sole focus of most research in this sub-field (Funke, 2005; Duckitt et al., 2010). In fact, some scholars even suggested that left-wing authoritarianism (LWA) was either exceptionally rare or, more often, non-existent (Stone, 1980; Stone & Smith, 1993; Altemeyer, 1996).

Recently, however, a growing number of political psychologists have been arguing more strongly for the existence of LWA and exploring its prevalence in various populations (Hiel et al., 2006; De Regt et al., 2011; Conway III et al., 2018; Fasce & Avendaño, 2020). Authoritarianism has been shown to be predicted by contagion threats (Murray et al., 2013) and to predict lower pro-environmental attitudes (Stanley & Wilson, 2019) as well as greater support for authoritarian political policies (Manson, 2020). Findings like these, suggesting that political polarization and forms of authoritarianism could decrease social cohesion and/or substantially influence public policy, are all the more crucial in the wake of climate change and the COVID-19 pandemic when social cohesion and flexible public policy is needed to address critical social issues.

In this article we present three studies that cumulatively provide insights into the character, cohesion, and clout of "tribal equalitarianism" (Clark & Winegard, 2020; Conway III, Zubrod, et al., 2020) in popular discourse, which we argue exhibits signatures suggestive of LWA and "binding" moral foundations of the sort normally associated with RWA. This tribal equalitarian discourse appears to be informed in large part by ideologies that fit broadly under the umbrella of "critical theory." Critical theory is an intellectual movement focused on critiquing social structures. Although its theoretical foundation was informed by ideas originally developed by Karl Marx, critical theory became popular in universities in the West following work by the "Frankfurt School" in the mid-1900s before the emergence and mainstreaming of post-modernism and social constructionism beginning in the 1960s.

Adorno, one of the key scholars in the assessment of political authoritarianism as right wing, also played a key role in the development of critical theory in academia, having co-authored one of its founding texts with Max Horkheimer (2002). In recent decades, however, some critical theorists have shifted toward a relatively more extreme ideological stance, colloquially associated with "woke" culture, which we suggest constitutes a tribal equalitarian discourse (Delgado and Stefancic, 2017). We analyze popular expressions of this tribal equalitarian discourse using data from Twitter and Google Books. The first four sub-sections below briefly introduce the relevant scientific literatures surrounding these issues and outline our rationale for and approach to studying tribal equalitarian discourses. The remaining three main sub-sections describe the methods and report results (and provide an initial discussion). We conclude with a general discussion of the implications of these findings, as well as the limitations of our research and the need for future work.

**Left-Wing Authoritarianism**

The existence and structure of LWA remains debated and understudied despite the fact that political movements with ideologies directly informed by Marx and his successors have been associated with over 100,000,000 deaths (Rummel, 1994). The existence and character of RWA, on the other hand, is well-documented and empirically validated. Indeed, many scholars continue to argue for asymmetry, maintaining Adorno's identification of authoritarianism with conservatism, an approach that is often guided by theories of motivated social cognition (Jost et al., 2003; Nilsson & Jost, 2020). Others have acknowledged that authoritarianism can indeed be found among some left-wing individuals, but argue that its presence among right-wing individuals is far more common and that, moreover, there are significant differences in the characteristics of RWA and LWA (Hiel et al., 2006). It has, for example, been observed that the tendency of the authoritarian left to withdraw from political engagement might be one of the factors leading to an apparent ideological asymmetry in the population (Federico et al., 2017).

Other scholars have defended the hypothesis of authoritarian symmetry – the idea that the characteristics and the rates of authoritarianism are similar among both the left and the right.
The symmetry hypothesis had its early champions (e.g., Ray, 1983), but has been growing in popularity in recent years as new evidence suggests that LWA does indeed exist, is correlated with political behavior and has characteristics that are quite similar to RWA (Conway III et al., 2018; Conway III & McFarland, 2019). This does not mean that LWA and RWA necessarily "mirror" one another, but rather that they share a "core" of traits such as "preference for social uniformity, prejudice toward different others, willingness to wield group authority to coerce behavior, cognitive rigidity, aggression and punitiveness towards perceived enemies, outsized concern for hierarchy, and moral absolutism" (Costello et al., 2020, p. 63).

One of the most studied factors influencing RWA is loss of (and need for) control. High levels of the latter can lead to compensatory attitudes and behaviors, including support for ideologies that help individuals regain and maintain a sense of control and structure (Kay et al., 2008). Such ideologies may be religious or secular, e.g., belief in a controlling god or in a controlling government (Kay et al., 2009). High need for control can also lead to ideological extremism (Kay & Eibach, 2013). In particular, people experiencing a loss of personal control are more likely to endorse a worldview composed of over-arching, totalizing, entities; and, as we unpack the particular beliefs related to the modern tribal egalitarian discourse in the next section, it is useful to keep in mind how central perceived *power imbalances* are to this worldview. Recent studies suggest that stress and threat may lead to increased conformity to the ideology of the status quo, whether conservative or liberal (Conway III, Chan, et al., 2020). Other studies show that liberals, like conservatives, conform more to ingroup norms in response to threatened personal control (Stollberg et al., 2017).

This does not mean that individuals high in RWA or LWA feel threatened in the same way nor that they favor the same solutions. Distinctive psychological needs and distinctive historical legacies play a role in the distribution of authoritarianism along left-right orientations (Thorisdottir et al., 2007; Dinas & Northmore-Ball, 2020). It also appears that liberals may be more responsive to *economic* policies as compensating for lost control, while conservatives are more responsive to *social* policies (Bullock et al., 2020; Ponce de Leon & Kay, 2020). This is supported by a study of a political activist sample in western Europe, which found that LWA was more negatively related to economic conservatism than cultural conservatism (Hiel et al., 2006). Research on left-authoritarians in western Europe also found that such individuals tended to privilege economic over socio-cultural concerns when joining a political party (Lefkofridi et al., 2014). Similarly, research on LWA in countries with communist legacies also shows that respondents in such contexts placed more emphasis on economic (over social) dimensions of their left-right orientation (Pop-Eleches & Tucker, 2020). This parallels the focus of communist regimes of the 20$^{th}$ century, who were, after all, premised on advocating for (coercive) economic redress.

It is also important to understand the debate over LWA-RWA (a)symmetry in the wider context of the ongoing debates over the extent to which liberals and conservatives are (un)equally biased (Ditto et al., 2019; Baron & Jost, 2019) and more or less (im)moral in their attitudes and behaviors (Smith et al., 2014; Vasilopoulos & Jost, 2020). This brings us to the literature surrounding another popular theory in social psychology—Moral Foundations Theory—that is relevant for our analysis of the presence and prevalence of LWA in tribal egalitarian discourse.

**Moral Foundations Theory**

Moral Foundations Theory (MFT) postulates the existence of (at least) five evolutionarily grounded intuitive "foundations" for morality (Graham et al., 2009; Haidt, 2007; Haidt & Joseph, 2004). Three of these – ingroup/loyalty, authority/respect, and purity/sanctity are commonly referred to as *binding* foundations because they are theorized to facilitate the cohesion of social coalitions. The remaining two – fairness/reciprocity and harm/care – are commonly referred to as *individualizing* foundations because of their apparent connection to the emphasis within the liberal philosophical tradition on the welfare and rights of individuals. The moral foundations questionnaire (MFQ) has been extensively utilized to show that while liberals tend to rely primarily on individualizing foundations, conservatives are more likely to rely on all five foundations. This reveals various combinations of preferred intuitive foundations or "moral signatures" (Haidt et al., 2009). The founders and most well-known proponents of MFT have explicitly framed the findings of their research not only in descriptive but also in normative terms, arguing that social scientists, who tend to be liberal, should place more value on the binding foundations of conservatives and embrace moral pluralism (e.g., Graham et al., 2013).

MFT is relevant for research on LWA and the broader issues related to the authoritarian (a)symmetry debate for at least two reasons. First, RWA has been shown to predict a "High Moralist" signature, which is characterized by undifferentiated high support across all five foundations (Milojev et al., 2014). This means that a typically "conservative" moral signature is one in which the binding foundations carry at least as much weight as the individualizing foundations. Such findings would seem to support the asymmetry thesis. However, precisely this sort of finding has led some scholars to criticize MFT in general and the MFQ in particular on theoretical and methodological grounds. For example, Hatemi and colleagues utilized panel data analysis to identify the causal direction between moral intuitions and political beliefs, and found that motivated reasoning driving the latter predicts the former rather than vice versa, challenging the explanatory power of MFT as a theory of ideology (Hatemi et al., 2019). MFT and the MFQ have also been criticized by proponents of the dual process model of ideology and prejudice, which argues that all morality is linked to harm in some way (Gray et al., 2014; Gray & Keeney, 2015; Schein & Gray, 2015), which helps to explain why the binding moral foundations sometimes fail to predict conservatism (Harnish et al., 2018).

Thus, in this paper we seek to discover how MFT might fare when utilized to find the "moral signature" emerging from public exchanges inspired or informed by critical theory discourse.

MFT is also relevant for our current research because of the role it has played in the wider discussion among social scientists about the *moral* (a)symmetry between conservatives and liberals. Are some combinations of the moral foundations "better" than others? Our purpose here is not to weigh in on this debate as moral philosophers but to contribute to the empirical evidence that might render such conversations more tractable. As noted above, the very idea that LWA might exist seems anathema to some social scientists. This reaction is not difficult to understand. For example, some studies indicate that the "individualizing" foundations (associated with the liberal left) have a universal *prejudice-reducing* effect, while the "binding" foundations (associated with the conservative right) *increase prejudice* against derogated outgroups perceived as threatening security and certainty (Hadarics & Kende, 2018). Prosocial orientations (such as empathic sensitivity) also seem to be more highly correlated with "individualizing" concerns (Strupp-Levitsky et al., 2020). Some evidence suggests that the Harm and Fairness scales of the MFQ reflect a *universalizing* motive (rather than the autonomy-seeking motive suggested by the MFT label "individualizing") whereas the Authority, Ingroup, and Purity scales reflect an *authoritarian* motive (Sinn & Hayes, 2017). These opposing effects on discrimination and intergroup hostility have led some critics to question the idea (common among proponents of MFT) that the binding and individualizing concerns should "be treated as operating on the same moral plane, objectively speaking" (Kugler et al., 2014, p. 426).

We agree with these (and other) authors that prejudice, outgroup hostility, etc., exacerbate rather than mitigate societal challenges, especially in contemporary pluralistic societies. However, it may well be that these and other "authoritarian" traits have a "binding" function among some left-leaning individuals.

**Tribal Equalitarianism**

Some of the worldview assumptions behind more authoritarian forms of left-wing political philosophy have recently been described as "tribal equalitarianism" (Conway III, Zubrod, et al., 2020). This worldview may appear paradoxical, or even oxymoronic, since "tribalism" (thinking one's own group is better than others) and "equalitarianism" (a tendency to minimize group differences) seem to be contradictory. Clark and Winegard have defined tribal equalitarianism as comprised of five interrelated components: (1) demographic groups (e.g., sexes, races) do *not* differ genetically with regard to socially valued traits; (2) prejudice and bigotry are *ubiquitous, constant and formative* in modern society; 3) the *only* reason groups differ is because of their regular experience of prejudice and discrimination; (4) anyone who asserts that groups differ for reasons other than prejudice or discrimination is only attempting to justify his or her own prejudices; and (5) we can and should make any arrangement to ensure that demographic groups in society enjoy equal outcomes" (2020, p. 10). It is certainly possible that a person embraces these equalitarian assumptions without being ideologically extreme. The point, however, is that these assumptions are in principle conducive to extreme, even violent retributive, policy attitudes due to construing reality as a war of power between groups.

Thus, the fact that fear at the extremes of *both* sides of the political ideological spectrum predicts negative emotion and outgroup derogation (van Prooijen et al., 2015) might lead us to expect to find individuals in some populations – under certain conditions – who are both left-wing *and* authoritarian. After all, evidence suggests that liberals (like conservatives) are more likely to operate out of, or endorse, the *binding* foundations when they have a high need for cognitive closure (Baldner et al., 2018) or high chronic locomotion concerns for controlling movement (Cornwell & Higgins, 2014). Evidence also suggests that Authority, Ingroup, and Purity foundations are at work in a left-liberal "emotive community" in the social sciences that sacralizes victims (of a certain sort) in a way that reduces academics' capacity to objectively appraise the circumstances of vulnerable groups (Horowitz et al., 2018).

As noted above, we are interested in studying the character, cohesion, and clout of a relatively extreme ideological trajectory, originally associated with critical theory in the academy, which has expanded into popular discourse in recent years. In what follows, we borrow the phrase "tribal equalitarian" to describe this discourse and explore the extent to which it exhibits traits related to the binding foundations of MFT and left-wing authoritarianism. The form of LWA on which we focus here has recently been described in a series of studies that identified a worldview (i.e., a set of variables that accounted for 55% of LWA's variance) that is "aggrieved and subversive," shaped by an "interpretation of the public sphere as deeply immoral, unfair, and uncertain," and which motivates people to "engage in extremist justice-seeking" (Fasce & Avendaño, 2020, p. 8).

The influence of academic critical theory on contemporary tribal equalitarian discourse and, potentially LWA, might be illustrated in hundreds of passages written across the decades. We'll provide two emblematic examples here. The first comes from a recent passage written by Angela Harris in her introduction to Richard Delgado's and Jean Stefancic's authoritative book *Critical Race Theory: An Introduction*:

> "Unlike traditional civil rights, which embraces incrementalism and step-by-step progress, critical race theory questions the very foundation of the liberal order, including equality theory, legal reasoning, Enlightenment rationalism, and neutral principles of constitutional law" (Delgado & Stefancic, 2017, p. 3).

As this quote demonstrates, critical theory's core axiom is that modern Western society is premised on prejudice and discrimination, as well as on institutional structures developed to support, or at least not challenge, the continuance of such prejudice and discrimination. For this reason, the sorts of polices, principles and institutions other civil rights leaders might depend on—equality theory, legal reasoning, scientific rationality, constitutional protections—are rendered moot. As such, effective redress comes from a revolutionary undoing and remaking of society and institutions to institute a new superstructure.

Capitalism is also a central source of critique for critical theorists and represents perhaps their core critique of inequality (a legacy of Marx's scholarship).

> *"Praxis begins with practice. This is the bedrock of revolutionary critical pedagogy's politics of solidarity and commitment. While radical scholarship and theoretical ideas are important—extremely important—people do not become politically aware and then take part in radical activity. Rather, participating in contentious acts of revolutionary struggle creates new protagonistic political identities that become refined through theoretical engagement and refreshed in every moment by practices of critical reflexivity. Critically informed political identities do not motivate revolutionary action but rather develop as a logical consequence of such action. And the action summoned by revolutionary critical educators is always heterogeneous, multifaceted, protagonistic, democratic and participatory—yet always focalized—anti-capitalist struggle,"* (McLaren, 2019, pg. 28).

Perhaps a more foundational example would better depict the same basic sentiment. Herbert Marcuse, a founder of critical theory in the 1960s in sociology writes:

"[The oppressed's] continued existence is more important than the preservation of abused rights and liberties which grant constitutional powers to those who oppress these minorities. It should be evident by now that the exercise of civil rights by those who don't have them presupposes the withdrawal of civil rights from those who prevent their exercise, and that liberation of the Damned of the Earth presupposes suppression not only of their old but also of their new masters," (Marcuse, 1965, p. 110).

Critical theory's core themes might be summarized as distrust of (Western) institutions, search and concern for inequality and oppression, and a desire to remove or reverse perceived status hierarchies. These would seem consonant with descriptions of the core of left-wing authoritarianism as "revolutionary aggression, top-down censorship, and anticonventionalism…[or] "morally absolutist and intolerant desires for coercive forms of social organization," (Costello et al., 2020, p. 70).

According to other innovations in tribal equalitarian discourse, individuals who belong to multiple categories that are viewed as oppressed (e.g., a black homosexual woman) are theorized to suffer from intersectional oppressions that compound the effect of being oppressed as a result of belonging to any single category (Crenshaw, 1989). While classical Marxist conflict theory focused primarily on economic delineations in power relationships, tribal equalitarian critical theory focuses more broadly on how exploitation is meted out across each social and personal identity held by oppressed individuals. This might be regarded as a theoretical extension of Marxist theory when blended with a focus on different constructed identities within the individual; justifying the inclusion of critical theory within the general framework of "Neo-Marxism" (Scott and Marshall, 2009). Tribal equalitarian discourse, consequently, often depicts the experience of oppression by beleaguered minorities with multiple maligned identities as constant, pervasive and mediated through every institution in society from family, economy and medicine to banking, law, and entertainment (Collins, 1990; Schulz & Mullings, 2006). Critical theorists often characterize social life as an all-encompassing matrix of intersecting oppressions (Bell, 1995; Collins, 2015; Delgado & Stefancic, 2017; DiAngelo, 2018; Ferber et al., 2007; Fleming, 2018; McIntosh, 1995; Sue, 2003, 2010) which are perpetuated—at least implicitly—by a legal and political superstructure that shapes the general social consciousness.

Prior analysts [e.g., (Pinker, 2005)] have noted that such discourses tend to conceptualize human behavior as being singularly driven by the pursuit of power: people do not interact with one another as individuals, rather, they interact as groups or "communities" motivated to seek greater power. This approach implies that the oppressed can only ensure their survival by dismantling privileged rights and institutional authorities, and remaking or re-organizing power relationships across institutions (Marcuse, 1965). Although tribal equalitarianism was initially conceived of by academics, it has subsequently spread into business, politics and popular culture. This is both a result of graduates from social sciences and humanities fields, wherein critical theory is taught , entering the workforce, but also a result of university administrators creating niches within universities that serve to promote and fundraise for programs discussing "diversity, equity and inclusion" which are, themselves, often on a foundation of critical theory (Campbell & Manning, 2018). Insofar as tribal equalitarianism frames reality as a struggle for power between groups deemed to have or "have-not" social and financial privilege and power, tribal equalitarian discourses could be a breeding ground for LWA.

**The Present Studies**

The following studies utilize social media analysis and bibliometric techniques (explained below) to explore the characteristics, coherence, and clout of tribal equalitarian discourses as manifested in the use of the cluster of terms that appear in Table 1.

*Table 1: Target Terms for Study*

| Terms | | | |
|---|---|---|---|
| Ableism | Genderfluid | Class oppression | sjw |
| Check your privilege | Genderqueer | Patriarchy | Stay woke |
| Cisgender | Healthy at any size | Prejudice plus power | Straightsplaining |
| cisgendered | Healthy at every size | Privilege | Thinsplaining |
| Cishet | Heteronormative | Privilege blindness | Third gender |
| Cisplaining | Heterosexism | Problematic | Toxic masculinity |
| Cissexism | Heterosplaining | problematize | Transphobia |
| Classplaining | Internalized oppression | Progressive stack | Trigger warning |
| Colonizers | Internalized inferiority | Preferred Pronouns | Triggering |
| Decolonize | Internalized superiority | Queer | Verbal violence |
| Cultural appropriation | Intersectionality | Rape culture | Victim blaming |
| Cultural erasure | Mansplaining | Safe space | White fragility |
| Gender erasure | Microaggression | Sizesplaining | White privilege |
| Racial erasure | Sexual oppression | Social construct | Whitesplaining |
| genderbinary | Racial oppression | Social justice | Woke |

The selection of these terms as representative of tribal equalitarian discourse in popular culture was the result of subject matter expert judgment, based on comprehensive reviews of the relevant academic literature and popular online discourse. Other experts might select other terms for analysis, but we believe that Table 1 captures our target adequately for our purposes. These terms might be considered as "shibboleths" because of how they might be used within this discourse to signal that one belongs to the tribe that holds equalitarian beliefs and attitudes to be sacred. The present studies were designed to answer hypotheses about the presence and prevalence of LWA sentiment and the "binding" foundations within tribal equalitarianism discourse, as well as the expansion of the latter as a dominant force in the culture wars raging in the U.S. and much of the English-speaking world.

**STUDY 1**

Although research on tribal equalitarian discourse and its possible relationship to LWA is growing rapidly, it has not yet been sufficiently established that such discourse contains signatures of ideological extremism and/or authoritarianism. Our first study aims to help fill this gap in the literature.

> H1: Tribal equalitarian discourse manifests signatures that are indicative of ideological extremism, authoritarianism, and the "binding" moral foundations.

*Methods*

To address this hypothesis, we studied data drawn from Twitter, which has 126 million daily active users and an average of more than 345,000 tweets per minute. Our data comes from a sample of posts scraped using the Twitter API between 4-18-2019 and 5-26-2019. Twitter was chosen because users can choose to associate keywords with their posts using the symbol "#," known as "hashtags." Tweets for the current study were scraped using hashtags that were created by removing any white spaces in phrases and adding the "#" symbol in front of any shibboleths (see Table 1 above) that appeared in the tweets. Using a script developed in Python, a total of 227,332 tweets were collected. Retweets (duplicates) were dropped from the analysis. In addition, to control for trolls in the data, we utilized an AI classification system trained to detect online "trolls" (created and provided by ALAN Analytics) and removed them from the dataset. This left 202,582 tweets, which included 7,353 unique tokens. Each tweet was then coded for moral signatures and term frequency analysis. In the results section below, these are referred to as Twitter dataset.

We also conducted a sentiment analysis on each specific term/hashtag in order to assess which of these were associated with the most emotion-laden language, and to determine whether the general emotionality of the tweets was consistent with established profiles of ideological *extremism* (Frimer et al., 2019). To test whether the Twitter dataset shows an *authoritarian* signature, we utilized Linguistic Inquiry and Word Count (LIWC) analysis (Lane, 2015; Lane & Gantley, 2018). Each of these analytical methods are discussed further below. Prior research utilizing LIWC has shown ideological extremism to be associated with consistently angry, negative tweets (Lane, 2015; Preoţiuc-Pietro et al., 2017). In addition, previous analyses using word frequency counts for key terms associated with different moral foundations (e.g., Graham et al., 2009) have indicated that political liberals typically use words that selectively emphasize individualizing or universalizing foundations (harm, fairness) while conservatives tend to use words related to the binding foundations (authority, ingroup, purity).

*Results and Discussion*

We found that the general emotionality expressed in the Twitter dataset is consistent with a profile of ideological extremism, i.e., words denoting anger/hostility are far more likely to be associated with the target terms/hashtags than are words denoting sadness or anxiety. The shibboleths found to be most associated with expressions of anger were: #cisgendered, #privilegeblindness, #rapeculture and #victimblaming, though other terms/hashtags also revealed this emotional signature (see Figure 1).

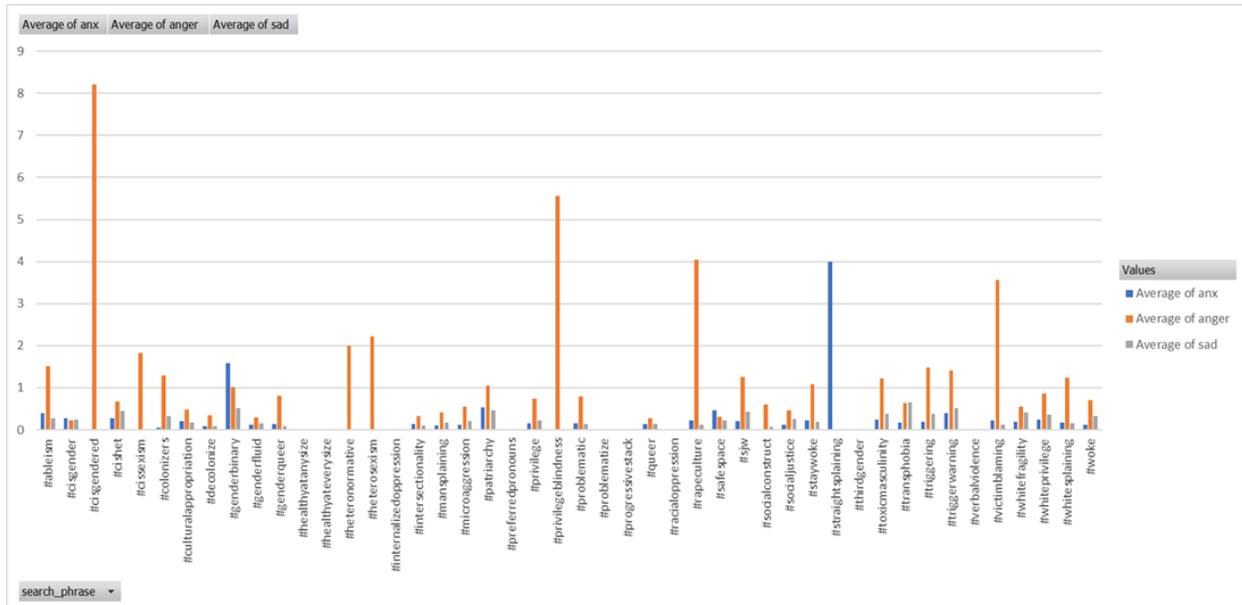

**Fig 1. Anxiety, Anger, and Sadness of DDW-Tweets**

Interestingly, only two shibboleths were prominently associated with expressions of anxiety: #straightsplaining (a term denoting heterosexual people attempting to explain things to others) and #genderbinary (a term denoting that male and female are the most common or only gender identities). Overall, expressions of sadness were far less common relative to expressions of anxiety and especially relative to the very common expression of anger. Of all emotions assessed, #cisgendered (a term denoting heterosexual people) was associated with the most emotion-laden text, almost all of it anger. Again, elevated levels of anger have been found in much prior work to be associated with political extremism and authoritarianism (e.g., Cohrs and Ibler, 2009; Strauts and Blanton, 2015; Sterling et al., 2020) perhaps because anger is selectively expressed in response to perceived moral transgressions (Gutierrez et al., 2012).

The results of our analysis of the moral foundations signatures of the text associated with each term/hashtag showed that texts indicative of moral foundations are most notably associated with #cisgendered, #healthyatanysize, #heterosexism, #privilegeblindness and #safespace. Figure 2 depicts the moral foundation domains associated with several terms/hashtags.

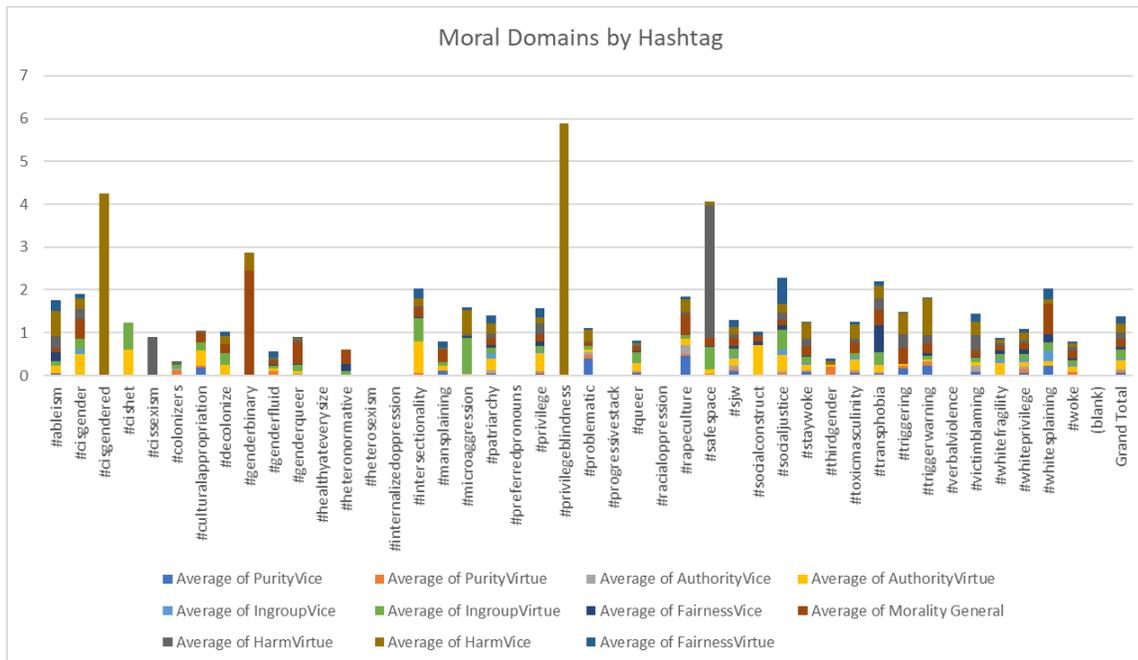

**Fig 2. Moral Domains by Hashtag**

We also assessed the overall moral foundations signature of text associated with all target terms/hashtags. This is important because, as noted above, prior research suggests that liberals and conservatives emphasize different sets of moral foundations: liberals tend to express the values of avoiding harm and ensuring fairness more than do conservatives, who place a greater relative emphasis on ingroup loyalty, deference to authority, and purity. We found that Twitter posts using our target terms contained an unusual moral signature not seen in prior research. These results indicate that this liberal discourse is, indeed, characterized by an emphasis on avoiding harm; however, it is also characterized by a greater emphasis on ingroup loyalty and deference to authority (binding foundations) than on fairness (see Figure 3c; Figures 3a and 3b show the idealized liberal and conservative moral signatures, respectively).

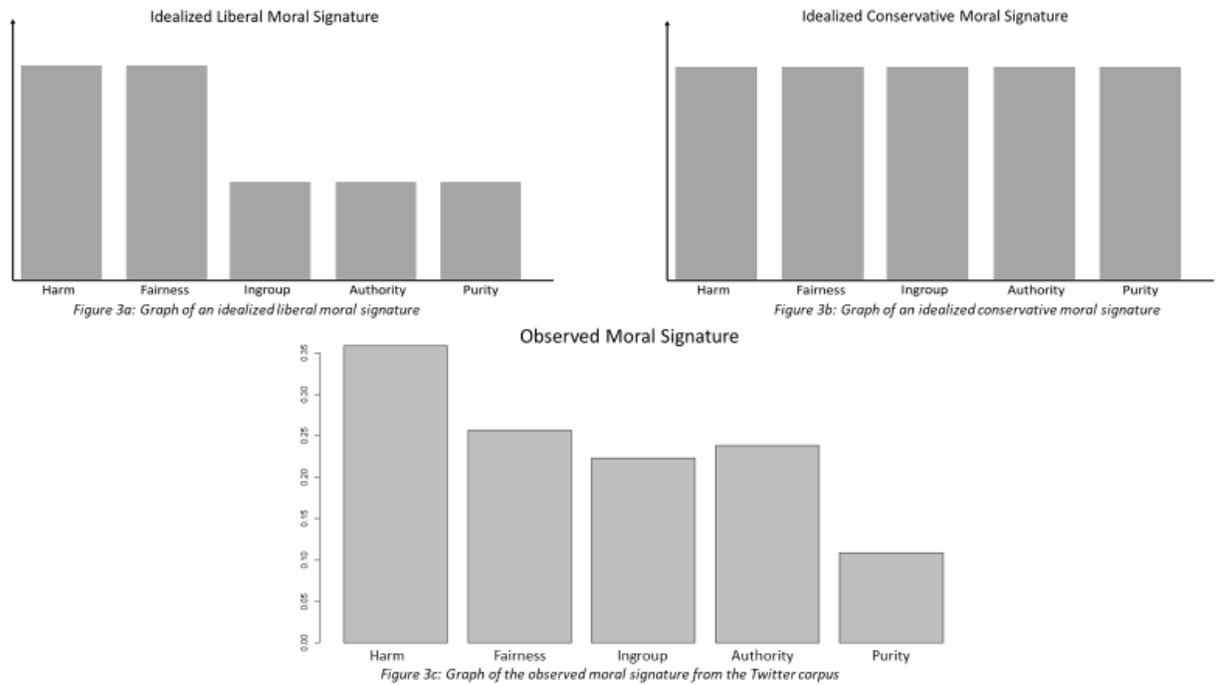

**Fig 3. Moral Domains of DDW-Tweets**

The discovery of this distinctive moral signature confirms other studies suggesting that there is more heterogeneity in ideological preferences than earlier MFT research suggested (Weber & Federico, 2013). This "tribal equalitarian" signature is highest on avoiding harm (characteristic of the liberal left), but higher on authority and ingroup foundations (characteristic of the conservative right) than it is on *fairness*. In other words, it is indeed "tribal" (focused on ingroup loyalty and deference to authority) but the tribe's interest in avoiding harming others may not actually be "fairly" (or equally) applied. What are we to make of the low score on "purity," a binding foundation associated with conservatism, in this moral signature? It could indicate that tribal equalitarian discourse is in fact low on concerns about purity/sanctity, weakening the claim that it is "tribal." Alternatively, and we think more likely, this could be an artifact of the focus in the MFQ on *sexual* purity concerns (Koleva et al., 2012), which are of less interest to equalitarians. The latter may well have scored higher on a scale that captured wider *sanctity* concerns, given their focus on sacred victims (Horowitz et al., 2018).

These findings could be interpreted as suggesting that the more angry and extreme a person's worldview becomes, regardless of whether that person identifies as liberal or conservative, the more likely he or she will emphasize ingroup loyalty and deference to (ingroup) authority. These traits could be part of a social coordination mechanism that supports conservatives in their attempts to preserve and protect existing traditions or institutions while supporting liberals in their attempts to deconstruct and reimagine existing traditions or institutions. These are questions for future research.

**STUDY 2**

Research on LWA has yet to establish whether or not discourse in the general public utilizing tribal equalitarian shibboleths forms a coherent discourse. Our second study aims to help fill this gap in the literature.

> H2: Tribal equalitarian discourse in the public sphere forms a complex, distributed, but cohesive semantic network.

*Methods*

To address our second hypothesis, we utilized 1) topic modeling, latent discourse analysis (LDA) and 2) semantic network analysis (Pennebaker et al., 2007; Lane & Gantley, 2018; Hill & Carley, 1999). In LDA, text from documents are used to create topics and build models of topics within and between documents. Topics are modeled as Dirichlet distributions. The analysis for this study was performed in Python using Gensim to extract topics from the Twitter dataset. Data from all hashtags were pooled into a single corpus for this analysis. To determine the appropriate number of topics for the LDA analysis, we ran a sweep of the k parameter, which controls the number of topics to be discerned from the data, from 1-56 (in steps of 6) to isolate where a possible increase (or "elbow") could be found that would suggest an optimal number for k. Within the intervals where statistically coherent topics could be found, we would sweep the k values within that interval in steps of 1 to isolate the primary k value for which the increase in coherence can be found. The coherence score analysis results are presented in the table in Appendix 1 of the online supplemental materials.

The semantic network analysis was conducted to investigate whether certain terms/concepts might be more central within the ideology than others, and whether there were multiple components within the network of concepts that were created by the corpus of tweets. Networks were constructed using a co-occurrence algorithm to create a network from the concepts in the text where each node represented a concept and links between nodes represented the co-occurrence of those concepts within a tweet (Lane, 2015). All tweets were analyzed to create a single network representing an abstraction of the general schema latent in the data.

*Results and Discussion*

The integrated topic model, inclusive of all target hashtags, provides support for the claim that our key term shibboleths (see Table 1) create a distributed worldview that is an amalgamation of many topics. Sweeping the k parameter up to 20 did not reveal any significant and acceptable variations in coherence scores (i.e., coherence score of .5 or above) using the elbow method for identifying an optimal k value. In order to better understand the data, further granularity was required, so models were created for k=20-50. While no distinct elbow could be found in the data, we did see a marked increase as k increased from 2 to 24, with a slight increase from 24-37, but no additional increases as the value of k increased from 37 to 50. For that reason, we chose 37 as the number of k values, as it was at that point where increases in coherence appear to level off, suggesting that increasing the number of topics discerned does not significantly improve the fit of the model. The LDA analysis can project the topics into two dimensions for visualization and further qualitative analysis (Figure 4).

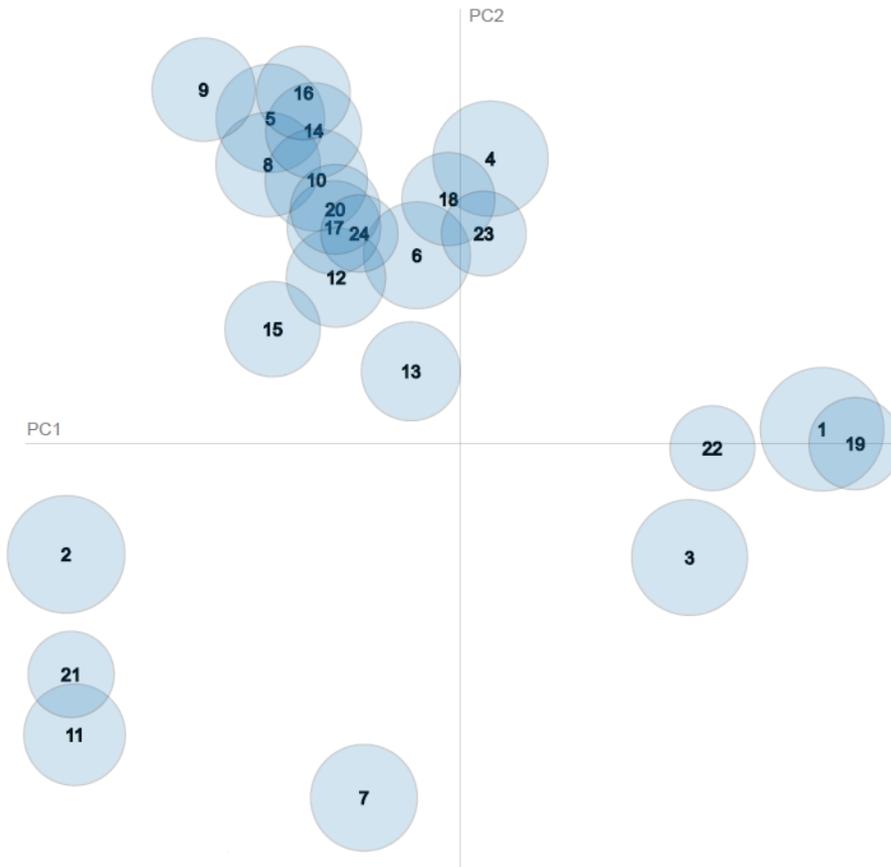

**Fig 4: Topic Model Visualization**

The projections in Figure 4 show that topics in this discourse are often overlapping, suggesting connections between the topics within the Twitter corpus. However, these topics are not completely clustered, indicating that some terms tend to be used more often in conjunction with one another. Despite this variation, two prominent clusters are visible in Figure 4. An online, interactive LDA mapping is provided in the github repository for this article. We also provide an interactive chart of the data in the electronic appendix, as well as a presentation of the most salient and relevant terms in the topic model.

Qualitatively, the cluster comprised of topics 1, 3, 19, and 22 (on the far right), appear to address topics related to economic inequality and include concepts such as "woman", "man", "black", "wage", "make," and "work". The cluster on the bottom, comprised only of topic 7, appears to focus on white privilege, as it includes topics such as "think", "know", and "privilege" as its most relevant terms but also includes the terms, "black", "safespace", and "work". The cluster to the far left, comprised of topics 2, 11, and 21, appear to address aspects of the term "wokeness". The most relevant terms in these topics include: "wake" and "stay" as well as "staywoke" (likely the result of stay and woke being combined as a hashtag), "law", "social justice", "sjw", "mansplaining", "rape", "life", "spread", "know", "report", and "campaign."

The large cluster in the top of the space includes all other topics discerned in the analysis. The terms within this cluster are generally varied if you consider the topics to be independent. However, the general inter-topic distance, depicted in the figure above, suggests that although they may be viewed as independent topics, they're not very semantically distant from one another. The key terms discerned through the analysis include terms as variant as whiteprivilege, woke, love, protest, Trump, diver (stemmed from diverse/diversity), gender, racist, activism, equality, sustainability, staywoke, transgend (stem of transgender), patriarchy, toxic, religion, white fragility, feminist, injustice, ableism, problematic, school, university, microaggression, trigger, queer, indentifi (stem of identify, identification), oppress, student, victim, decolonize, whitesplain, discrimination. The top terms in this cluster of topics can all be seen in the interactive visualization in the appendix.

However, while it appears as if there are 4 clusters of topics in the PCA visualization above (Figure 4), when creating a topic model of k=4 (in an attempt to see if those clusters can collapse into fewer, larger, clusters), the coherence score was .3564, which is lower than the acceptable threshold (i.e.,values being >.5) and lower than the chosen value of k=37, discussed above, which did provide an acceptable cohesion score. To explain this, future research may wish to look further into how this worldview may be socio-semantically clustered; i.e., it may be that statistically discernable sub-groups within the overall network stress certain topics or ideas to a greater extent than others.

As noted above, we also utilized an additional test to help assess if this worldview can be called properly called cohesive. Specifically, we used semantic network mapping to analyze the interrelationships between concepts in the Twitter dataset. We found that although many terms, such as screennames, were isolated in the network, the network as a whole had one clearly defined giant component (see Figure 5).

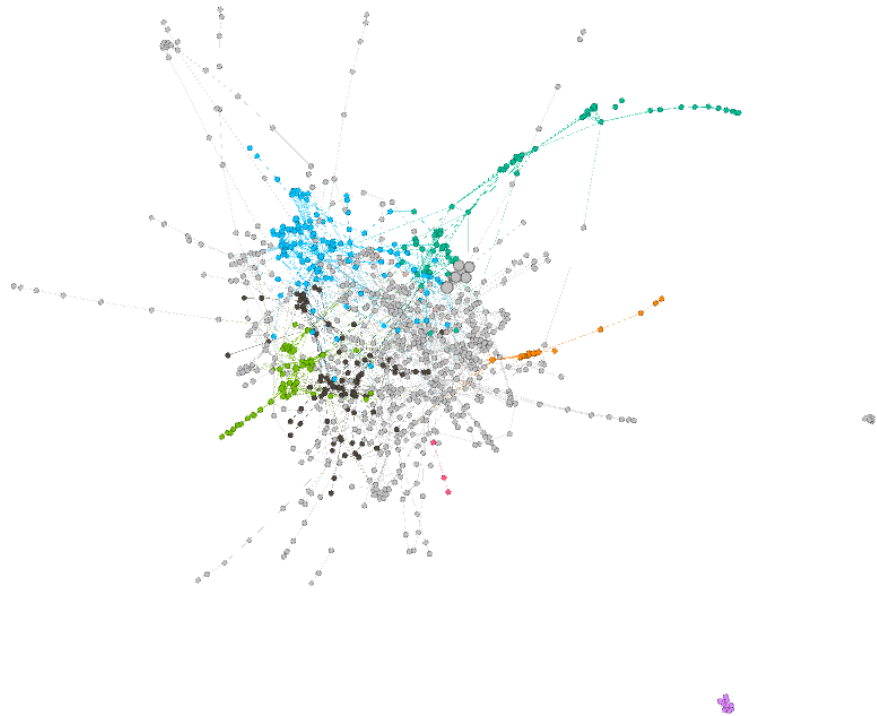

**Fig 5. Visualization of Giant Component.**
Colors represent Louvain modularity classes, and the sizes are determined by weighted degree centrality.

After filtering isolate nodes that we took to be noise in the network, the network was comprised of 935 nodes and 2676 links. We then analyzed all nodes in the network for their network centrality. The higher clustering coefficient of the observed network compared to an Erdös-Rényi graph of similar construction would suggest that the network observed here is a scale-free small-world network. This network is reminiscent of other networks that exist in the real world and similar to those drawn from a single corpus of materials (Lane, 2015; Ravasz & Barabási, 2003; Steyvers & Tenenbaum, 2005). Additionally, using the Humphreys & Gurney quantitative definition of a scale-free small-world network, given the observed network properties, and a random network of similar construction, we find a $S\Delta > 10$, which also suggests that the network is a scale-free small world network. The network also exhibits a clustering coefficient (and $S^\Delta > 10$) that is far greater than expected and aligns with similar network properties discerned from semantic networks created from large corpora of religious texts (Lane, 2015).

Results of calculating the degree, eigencentrality, and triangles for each term showed that there is generally a stable "core" of beliefs that exist across the different topical hashtags used to collect the data. The top 10 terms for each of the three measures are presented in Table 2 below.

| Id | Degree | Id | Eigencentrality | Id | Triangles |
|---|---|---|---|---|---|
| queer | 63 | queer | 1 | queer | 141 |
| woke | 44 | woke | 0.713502 | peopl | 103 |
| like | 42 | like | 0.618987 | like | 99 |
| peopl | 36 | lgbtq | 0.564906 | lgbtq | 92 |
| lgbtq | 35 | got | 0.560448 | woke | 83 |
| new | 33 | get | 0.531244 | know | 83 |
| get | 33 | peopl | 0.510014 | system | 81 |
| system | 32 | know | 0.499587 | new | 77 |
| got | 29 | system | 0.472198 | got | 75 |
| men | 28 | new | 0.464294 | indigen | 63 |

**Table 2: Term Total Degree Centrality**

One of the limitations of these studies is that Twitter is not representative of the American public: Twitter users are younger, more formally educated and have higher incomes than the general American public (Wojcik & Hughes, 2019). Twitter users are also more likely to identify as Democrat (60% of Twitter users compared to 52% of the general population) and less likely to identify as Republican (35% of Twitter users compared to 43% of the general public). However, the left-leaning political skew of Twitter is arguably an appropriate population of people from which to sample, insofar as an authoritarian strain of tribal equalitarian discourse, informed by critical theory, is more common among liberals.

## STUDY 3

Finally, the extent to which critical theory-informed tribal equalitarian discourse has grown in influence in the academy and the public sphere has yet to be established. Our third study aims to fill this gap in the literature.

> H3: The prevalence of critical theory-informed tribal equalitarian discourse has expanded rapidly and substantively in public discourse in recent years.

*Method*

To address hypothesis 3, we searched through the English Corpus of Google Books, available through their "Ngram" system. The 2019 English corpus was released in 2020, which updated the previous corpus of over 1 trillion words and 8 million books with additional books and more advanced optical character recognition. Our rationale was that detecting widespread distribution of terms derived from critical theory works in publications such as books over the past few decades would indicate their increased acceptance and clout in the public sphere.

As a lens into human culture, this dataset is very unique because it holds a great deal of historical information about the relative frequencies of different phrases. Since data is presented longitudinally, we are better able to understand the relative changes in term usage over time from published media. It is important to note that this data is not without sampling bias itself. Because it only focuses on published books, it neglects a great deal of other media and therefore reflects a bias toward term usage and beliefs prevalent among scholars and educated lay readers. However, this bias is acceptable for the purposes of the current study because, as noted above, critical theory was initiated within the academy but seems to have expanded, in the form of tribal equalitarianism, into public discourse.

*Results and Discussion*

Our results provide support for hypothesis 3. As shown in Figure 6 below, in which all search terms are combined into a single query, we see a marked increase since 1960, leveling off to a plateau around the year 2000, but then continuing with a sharp uptick beginning in 2009 and continuing to the present.

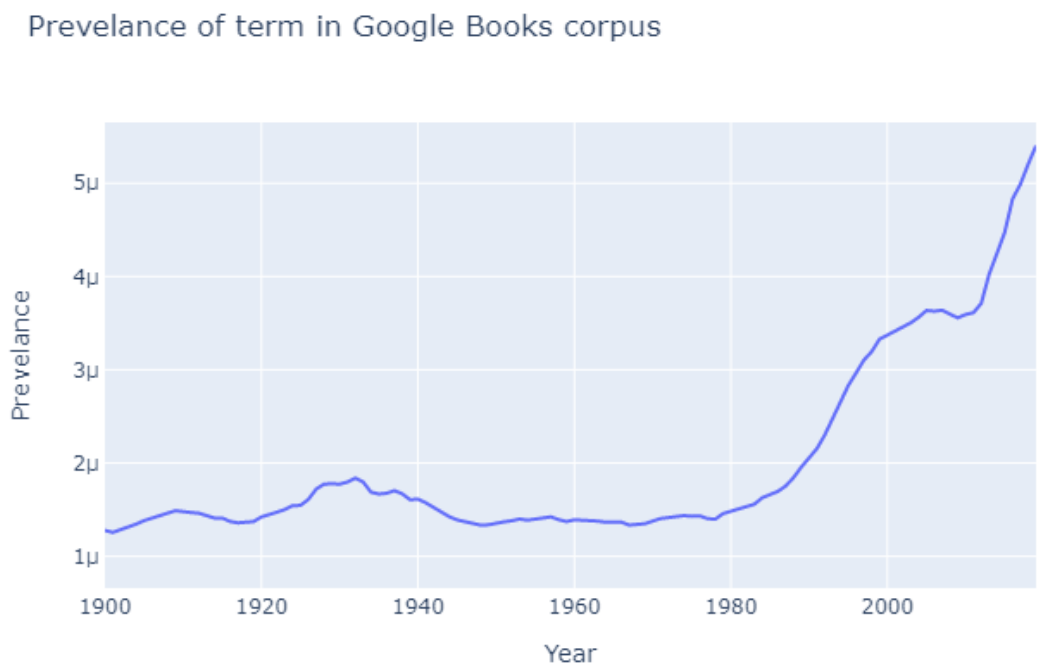

*Fig 6. Combined Prevalence of Target Terms*

The fact that the use of key target terms increased rapidly at the beginning of the 2008-2009 financial collapse could be interpreted in light of research suggesting that ideologies may be compensatory and made more salient during periods of uncertainty or helplessness. The initial post-1980 trend also coincides with two important economic shifts: 1) a marked period of the growth of wealth inequality in the English speaking world, where in the US, for example, estimates of inequality show an increase of about 20% from 1980-2016 (Elwell et al., 2019) and 2) the decline, starting in the late 1970s and early 1980s, of well paid, permanent, tenured faculty at American universities (Weissmann, 2013) happening in conjunction with the number of PhDs awarded—rising from 33,497 in 1988 to 55,195 in 2018. While the first shift is more obviously related to the overall economic context within which these texts were written, the second shift should not be discounted offhand as it would represent an immediate perception in the loss of power for a great number of the individuals who were producing the professional texts that are known to bias the period of the Google Books English Corpus covered here (Pechenick et al., 2015).

We also investigated the rise of different term shibboleths by analyzing their growth patterns individually, which revealed several noteworthy patterns. Visualizations of all data retrieved from 1800-2019 are available in the online appendix. One such pattern is seen in the use of several terms, such as "white privilege," which show a marked increase starting in the late 90s and increasing rapidly until the end of the dataset. This pattern is observed in other terms as well, such as "social construct" as well as "safe space," "transphobia," "internalized oppression", "ableism," "woke", "genderqueer," "heteronormative," "decolonize," "rape culture," "queer", "third gender", "victim blaming", and "microaggression."

However, the most pronounced incline found in our dataset comes from the term "intersectionality," which was not found in the dataset until 1984, and then increased rapidly until 2019, the end of the available data (see Figure 7).

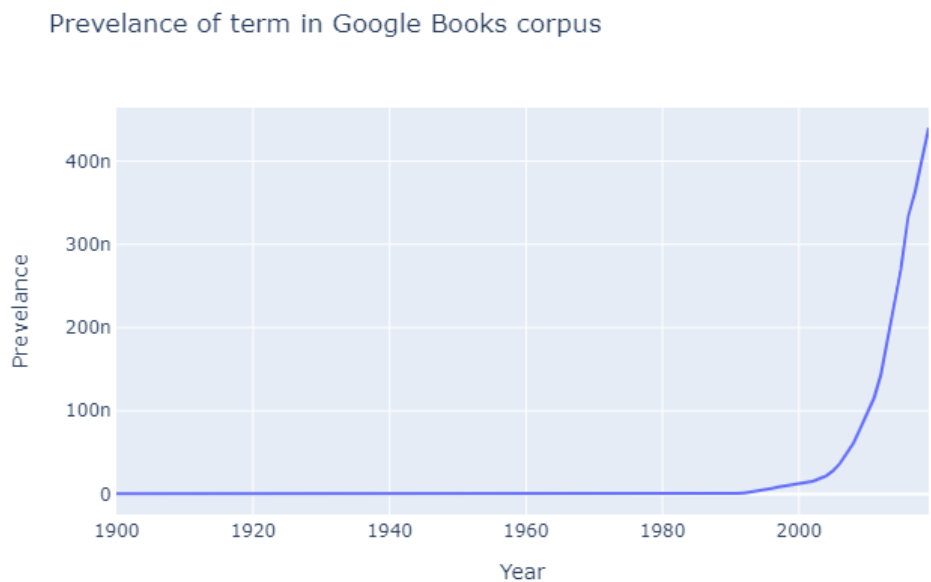

*Fig 7. Intersectionality*

Another set of terms show a general trend of growth with the added caveat that their usage was already steady prior to the 1950s. These terms include "social justice," "dialectic," "queer," "privilege," "triggering" and "colonizers." For additional figures and analysis, see the online supplemental materials (LINK TO SUPPLEMENTAL MATERIALS REDDACTED FOR REVIEW CURRENTLY STORED IN PRIVATE GITHUB ACCOUNT TO BE MADE PUBLIC UPON PUBLICATION).

**General Discussion**

Before sharing some concluding thoughts, we'd like to underscore some important limitation of this research. First, Twitter is not representative of the American public: Twitter users are younger, more formally educated and have higher incomes than the general American public (Pew, 2019). Twitter users are also more likely to identify as Democrat (60% of Twitter users compared to 52% of the general population) and less likely to identify as Republican (35% of Twitter users compared to 43% of the general public). However, the left-leaning political skew of Twitter is arguably an appropriate population of people to sample from, insofar as the ideology described above is particularly attractive to liberals.

A second limitation relates to the time frame wherein tweets were scraped. Though we are not aware of any unusual political event that occurred during these dates that might skew our findings, it is nevertheless possible that window of time during which data was collected may have included events that affected results. If our analysis is purely the result of a brief period-effect, however, it is surprising that the worldview nevertheless cohered in a way that was consistent with how critical theory, and its more popularized tribal equalitarianism, is discussed in relevant academic and popular culture sources. Due to this consistency, we think it is implausible that our findings are merely the result of the time frame within which we chose to scrape tweets.

Our central finding was that, with the exception of the moral foundation of purity (which is over-determined in its measurement by relying on an assessment of peoples' concern with sexual purity), text associated with the target terms/hashtags did indeed reveal an emphasis on ingroup loyalty, authority and anger. This suggests the possibility that, regardless of whether a person identifies as liberal or conservative, the more angry and extreme a worldview becomes, the greater the emphasis on ingroup loyalty and deference to (ingroup) authority. Perhaps this loyalty and deference to authority is part of a social coordination mechanism which, for conservatives, aids in preserving and protecting existing traditions or institutions and, for liberals, aids in dismantling and reimagining traditions or institutions. These are key questions for future research.

Some important caveats should also be mentioned. Our argument here has not been about the truth or tenability of academic critical theory or of its popularization in tribal equalitarian discourse. We share the concerns expressed by the latter about the de facto existence of the discrimination, bigotry and inequality that cause so many problems in contemporary societies. We also acknowledge that authoritarianism is also found on the far political right, where one all too often finds denigrating attitudes and behaviors toward the poor, women, racial minorities and homosexuals (Osborn and Weiner, 2015; Tope et al., 2015; Hodson and MacInnis, 2017; van der Toorn et al., 2017; Smith et al., 2019). As noted above, there has been far more scholarship on RWA and the political right (e.g., Whitely, 1999; Zakrisson, 2005; Duckitt et al., 2010) than LWA and the political left, and our goal has been to contribute to the scientific literature on the latter.

We have also tried to contribute toward a methodological pluralism in scholarship on authoritarianism. In addition to continuing to assess the particular ideological, attitudinal or conceptual correlates of both right-wing and left-wing authoritarianism, we encourage researchers to begin studying such correlates in more "natural" samples outside of experimental settings and outside of surveys. Much of our existing understanding of authoritarianism, right and left, comes from surveys; while informative, the latter can be complemented by (anonymized) analyses of internet media. Approaches using "big data" can also begin to assess "cultural evolutionary" dynamics that would be harder to assess using surveys. For example, some scholars have suggested that more extreme policy positions taken by activists on the left (or right) might trigger backlash responses represented in an extreme policy position advocated by those on the right (or left), which leads to a doubling down on more extreme positions on the left (or right) in response and so on (Campbell and Manning, 2018). This sort of "cultural evolution" of worldviews toward more extreme positions might be akin to processes of inter-group polarization where the most vocal critics of each party voice their frustrations, moving the group norm closer and closer over time to more extreme positions (Sunstein, 1999).

Future work could also use additional survey or secondary big data research to further test the hypothesis that the adoption of critical theory in the academy is a crucial factor in the adoption of tribal equalitarianism in the public discourse. Intermediary sources of cultural information, namely the print media, could also be used to study the rise in tribal equalitarian discourse, as the shift from academia to popular discourse is likely both a bottom up (e.g., many graduates having been taught these theoretical perspectives in universities) and top down (e.g., editors in major newspapers perpetuating the ideological signatures of the ideology) cultural process (regarding major news media, see Goldberg, 2020).Further work might also investigate how the perception of different environmental threats affect moral foundations, political alignments, and authoritarianisms. This research might explicate why some social or environmental threats result in a wider (i.e., across the political spectrum) adoption of authoritarianism.

Finally, the spread of LWA/RWA in online social networks should be further investigated, as it appears that the emotional-moral nature of these politically charged topics encourages engagement in online social media, which may then be perpetuated in a feedback loop as "deep-learning" algorithms used on platforms such as Twitter encourage further engagement (Koumchatzky and Andryeyev, 2017). Such an effect could 1) bias the information presented to individuals in favor of LWA/RWA; 2) echo chambers could begin to occur on Twitter around these issues (both for and against their ideas); 3) activate natural biases, such as the information availability bias could lead people to believe that LWA/RWA is more widespread than it actually is; or 4) a mixture of 1 and 3, whereby authoritarianism grows in dangerous ways because LWA and RWA are exacerbated algorithmically by "deep learning" social media technologies, leading to a dissolution of cohesion and associated democratic processes in times when cohesion and thoughtful political discourse are most needed.

Ditto, P. H., Liu, B. S., Clark, C. J., Wojcik, S. P., Chen, E. E., Grady, R. H., Celniker, J. B., & Zinger, J. F. (2019). At least bias is bipartisan: A meta-analytic comparison of partisan bias in liberals and conservatives. *Perspectives on Psychological Science*, *14*(2), 273–291.

Duckitt, J., Bizumic, B., Krauss, S. W., & Heled, E. (2010). A tripartite approach to right-wing authoritarianism: The authoritarianism-conservatism-traditionalism model. *Political Psychology*, *31*(5), 685–715.

Durkheim E. The elementary forms of religious life. ([1912]1965) New York: The Free Press.

Elwell, J., Corinth, K., & Burkhauser, R. V. (2019). *Income Growth and its Distribution from Eisenhower to Obama: The Growing Importance of In-Kind Transfers (1959-2016)* (No. w26439). National Bureau of Economic Research. https://doi.org/10.3386/w26439

Fasce, A., & Avendaño, D. (2020). Opening the can of worms: A comprehensive examination of authoritarianism. *Personality and Individual Differences*, *163*, 110057.

Federico, C. M., Fisher, E. L., & Deason, G. (2017). The authoritarian left withdraws from politics: Ideological asymmetry in the relationship between authoritarianism and political engagement. *The Journal of Politics*, *79*(3), 1010–1023.

Ferber, A. L., Herrera, A. O., & Samuels, D. R. (2007). The matrix of oppression and privilege: Theory and practice for the new millennium. *American Behavioral Scientist*, *51*(4), 516–531.

Finkel, E. J., Bail, C. A., Cikara, M., Ditto, P. H., Iyengar, S., Klar, S., Mason, L., McGrath, M. C., Nyhan, B., & Rand, D. G. (2020). Political Sectarianism: A Dangerous Cocktail of Othering, Aversion, and Moralization. *Science*, *370*(6516), 533–536.25

Haidt, J., Graham, J., & Joseph, C. (2009). Above and below left–right: Ideological narratives and moral foundations. *Psychological Inquiry*, *20*(2–3), 110–119.

Haidt, J., & Joseph, C. (2004). Intuitive ethics: How innately prepared intuitions generate culturally variable virtues. *Daedalus*, *133*(4), 55–66.

Harnish, R. J., Bridges, K. R., & Gump, J. T. (2018). Predicting economic, social, and foreign policy conservatism: The role of right-wing authoritarianism, social dominance orientation, moral foundations orientation, and religious fundamentalism. *Current Psychology*, *37*(3), 668–679.

Hatemi, P. K., Crabtree, C., & Smith, K. B. (2019). Ideology Justifies Morality: Political Beliefs Predict Moral Foundations. *American Journal of Political Science*, *63*(4), 788–806. https://doi.org/10.1111/ajps.12448

Hiel, A. V., Duriez, B., & Kossowska, M. (2006). The Presence of Left-Wing Authoritarianism in Western Europe and Its Relationship with Conservative Ideology. *Political Psychology*, *27*(5), 769–793. https://doi.org/10.1111/j.1467-9221.2006.00532.x

Hill, V., & Carley, K. M. (1999). An approach to identifying consensus in a subfield: The case of organizational culture. *Poetics*, *27*(1), 1–30. https://doi.org/10.1016/S0304-422X(99)00004-2

Hodson G, MacInnis CC. (2017) Can left-right differences in abortion support be explained by sexism? Pers Individ Dif. 104:118-21.

Horkheimer, M., Adorno, T. W., & Noeri, G. (2002). *Dialectic of enlightenment*. Stanford University Press.

Horowitz, M., Haynor, A., & Kickham, K. (2018). Sociology's sacred victims and the politics of knowledge: Moral foundations theory and disciplinary controversies. *The American Sociologist*, *49*(4), 459–495.
27

Pechenick, E. A., Danforth, C. M., & Dodds, P. S. (2015). Characterizing the Google Books Corpus: Strong Limits to Inferences of Socio-Cultural and Linguistic Evolution. *PLOS ONE*, *10*(10), e0137041. https://doi.org/10.1371/journal.pone.0137041

Pennebaker, J. W., Booth, R. J., & Francis, M. E. (2007). *Linguistic inquiry and word count*. LIWC. liwc.net

Pew Research Center. Sizing up Twitter users. Accessed 10/19: https://www.pewinternet.org/2019/04/24/sizing-up-twitter-users/

Pinker, S. (2005). *The blank slate*. Southern Utah University.

Ponce de Leon, R., & Kay, A. C. (2020). Political ideology and compensatory control mechanisms. *Current Opinion in Behavioral Sciences*, *34*, 112–117. https://doi.org/10.1016/j.cobeha.2020.02.013

Pop-Eleches, G., & Tucker, J. A. (2020). Communist legacies and left-authoritarianism. *Comparative Political Studies*, *53*(12), 1861–1889.

Preoţiuc-Pietro, D., Liu, Y., Hopkins, D., & Ungar, L. (2017). Beyond binary labels: Political ideology prediction of twitter users. *Proceedings of the 55th Annual Meeting of the Association for Computational Linguistics (Volume 1: Long Papers)*, 729–740.

Ravasz, E., & Barabási, A.-L. (2003). Hierarchical organization in complex networks. *Physical Review. E, Statistical, Nonlinear, and Soft Matter Physics*, *67*(2), 026112-1-026112–026117. https://doi.org/10.1103/PhysRevE.67.026112

Ray, J. J. (1983). Half of all authoritarians are left wing: A reply to Eysenck and Stone. *Political Psychology*, 139–143.

Rummel, RJ (1994) *Death by Government: Genocide and Mass Murder in the Twentieth Century*. New Brunswick, NJ: Transaction
30